\documentclass[aps,preprint,amsmath,amssymb,groupedaddress,preprint]{revtex4-2}
\usepackage{amsmath}
\usepackage{graphicx}
\usepackage{color}
\usepackage{color}
\usepackage{graphicx}
\usepackage{natbib}
\usepackage{epsfig}
\usepackage{setspace}
\usepackage{amsmath,amssymb}
\usepackage{verbatim}
\usepackage{bibentry}
\usepackage{xcolor}
\usepackage{bm}
\usepackage{lipsum}

\begin{document}

	\title{Thermoelectric performance of quantum dots embedded in an Aharonov-Bohm ring: a Pauli master equation approach}
	
	\author{Parbati Senapati}
	\affiliation{Department of Physics,
		Indian Institute of Technology Patna, Bihta, Bihar, 801106, India}
	\author{Prakash Parida}\email{pparida@iitp.ac.in}
	\affiliation{Department of Physics,
		Indian Institute of Technology Patna, Bihta, Bihar, 801106, India}

%\keywords{Keyword1, Keyword2, Keyword3}

\begin{abstract}
	Within linear response theory using Pauli master equation approach, we have investigated the thermoelectric properties of quantum dots (QDs) embedded in an Aharonov-Bohm (AB) ring weakly coupled to two metallic electrodes. This study explores the impact of magnetic flux on thermoelectric transport, emphasizing the role of quantum interference induced by the flux. When the magnetic flux is varied from 0 to one quantum of flux \(\Phi = \Phi_{0} = \frac{h}{e}\), both the electrical conductance and the thermoelectric figure of merit (\(ZT\)) significantly increase by two order of magnitude. Moreover, our investigation into the effects of onsite and inter-site Coulomb interactions in this nanojunction indicates that an optimal $ZT$ is attained with moderate onsite Coulomb interaction and minimal inter-site Coulomb interaction. We briefly discussed the effects of asymmetric arrangements of triple QDs within an AB ring. However, within our parameter regime, a symmetric arrangement offers superior thermoelectric performance compared to asymmetric configurations. Furthermore, we explored how increasing the number of QDs in the ring enhances the thermoelectric properties, resulting in a potential $ZT$ value of around $0.43$. This study shows that arranging multiple QDs symmetrically in an AB ring can result in significant thermoelectric performance in nanostructured system at low temperatures.
\end{abstract}

\maketitle

\section{Introduction}
In the relentless pursuit of energy harvesting devices adept at converting waste heat into electricity, recent decades have witnessed an intense quest. The thermoelectric phenomenon, enabling the conversion between thermal and electrical energies, has attracted substantial theoretical and experimental interest in recent years \cite{1,2,3,dubi2011colloquium,boukai2008silicon,wierzbicki2010electric,beenakker1992theory,benenti2017fundamental,kuo2024thermoelectric}. Despite considerable interest, the widespread adoption of thermoelectric energy conversion technology has been constrained, mainly due to its relatively low efficiency. Although specialized applications, such as laboratory equipment and space missions, have proven its viability, broader use remains limited by the need for significant performance improvements. The thermoelectric performance is typically characterized by the dimensionless figure of merit $ZT=\frac{S^{2}G_{V}T}{k_{e}+k_{ph}}$, where $T$ represents the temperature of the system. The thermopower, also known as the Seebeck coefficient ($S$), quantifies the magnitude of the voltage induced in response to a given temperature gradient with no current. Meanwhile, the electrical conductance ($G_V$) reflects the ease with which charges can flow through the system to generate a voltage drop. Additionally, electron thermal conductance ($k_e$) and phonon thermal conductance ($k_{ph}$) gauge the resistance to heat transfer across the system, thereby maintaining a temperature gradient \cite{dubi2011colloquium}.

To enhance the magnitude of $ZT$, it necessitates a combination of high thermopower, high electrical conductivity, and low thermal conductivity. Nonetheless, in conventional bulk thermoelectric materials, these physical quantities are interdependent, as dictated by the Wiedemann-Franz law \cite{dusastre2010materials,wilson2012experimental}, where the ratio between electrical and thermal conductivities remains constant \cite{snyder2008complex,hicks1993thermoelectric}. Consequently, an increase in electrical conductance typically results in a corresponding rise in thermal conductance and is consistently accompanied by a reduction in thermopower. However, for widespread commercial application, thermoelectric materials with even greater $ZT$ values are necessary. To this end, various approaches have been proposed to enhance thermoelectric performance, including reduction of system dimensionality \cite{wang2013enhancement,karamitaheri2012engineering,sothmann2014thermoelectric}. Consequently, intensive research has focused on the thermoelectric effect in zero-dimensional quantum dots (QDs), particularly in the Coulomb blockade \cite{kubala2008violation,dubi2009thermospin,swirkowicz2009thermoelectric,liu2010enhancement,parida2009negative} and Kondo regimes \cite{scheibner2005thermopower,dong2002effect}. The significant enhancement in thermoelectric performance in such devices is attributed to a pronounced deviation from the Wiedemann-Franz law driven by Coulomb interactions, along with a substantial reduction in thermal conductivity ($k_{ph}$) due to intense phonon scattering at the interfaces of nanostructures \cite{kuo2010thermoelectric}.

However, investigative efforts are also focused on comprehending the significant influence of symmetries, phase coherence, external magnetic fields, and quantum interference phenomena \cite{buvca2012note,manzano2018harnessing,rai2011magnetic,engelhardt2019tuning,thingna2020magnetic,thingna2016dynamical,das2023asymmetric}. The renowned Aharonov-Bohm (AB) interferometer serves as a prime example of a phase-tunable quantum device \cite{aharonov1959significance,yeyati1995aharonov,kotimaki2010aharonov,haack2019efficient}. This phenomenon manipulates electron wave function phases within the nanostructure through magnetic field threading, leading to profound interference effects that elucidate the fundamental quantum nature of charge and thermoelectric transport in mesoscopic systems.
Numerous studies on thermoelectric properties of QDs AB interferometers have been conducted by various researchers, consistently revealing favorable thermoelectric characteristics \cite{gores2000fano,holleitner2001coherent,liu2007transport}. Liu et al. investigated intra-dot Coulomb interaction and quantum interference impact on thermoelectric properties in an AB interferometer with interacting QDs, suggesting potential for high-efficiency solid-state thermoelectric conversion \cite{liu2011role}. Zheng et al. explored the thermoelectric effect in an AB interferometer with an embedded QD using non-equilibrium green's function (NEGF) formalism, revealing a remarkably high ZT value attributed to the Fano effect from quantum interference \cite{zheng2012thermoelectric}. Behera et al. also utilized NEGF formalism for thermoelectric studies and found that the symmetric triple-QD AB interferometer allows better thermoelectric response control than its asymmetric counterpart, functioning as a thermoelectric heat engine \cite{behera2023quantum}. Wang et al. provided a comprehensive summary of both theoretical and experimental advancements in inelastic thermoelectric transport and fluctuations in mesoscopic systems. Their study revealed novel phenomena arising from inelastic thermoelectric transport, including linear thermal transistors, cooling by heating, heat-charge cross rectification, and cooling driven by thermal currents \cite{wang2022inelastic}. Lu et al. developed a theory for optimal efficiency and power in three-terminal thermoelectric engines with two independent output electric currents and one input heat current. Their numerical calculations for a triple-QD thermoelectric engine show significant improvements in efficiency and power when using two output electric currents \cite{lu2019optimal}.
\par
All above theoretical thermoelectric studies of QDs embedded in an AB interferometers have predominantly utilized the NEGF formalism as well as the Landauer-Büttiker approach. However, our work focuses on the weak coupling regime and leverage the Pauli master equation (PME) \cite{muralidharan2012performance,senapati2024charge} to explore thermoelectric transport properties via sequential electron tunneling. Although NEGF formalism is a powerful tool for quantum transport studies, it faces challenges in efficiently managing many-body interactions, especially in systems with strong Coulomb correlations or complex many-body effects. Similarly, the Landauer-B\"{u}ttiker formula is a well-established method for calculating transport properties in semiconductor nanostructures within the steady-state quantum transport regime. It fundamentally links the electron wave functions (scattering amplitudes) of a quantum system to its conductance properties, expressing transport current in terms of transmission coefficients derived from the single-particle scattering matrix. However, this approach assumes that the reservoirs connected to the scatterer (mesoscopic system) remain in equilibrium and that electrons within these reservoirs are always incoherent. As a result, the Landauer-B\"{u}ttiker formula is not applicable to transient quantum transport, where nonequilibrium effects and time-dependent dynamics play a crucial role \cite{yang2017master,yang2015master}. In contrast, the master equation approach offers a comprehensive framework for studying quantum transport by using the reduced density matrix to capture quantum coherence, decoherence, and non-Markovian effects, making it valid for both steady-state and transient dynamics. Additionally, the master equation approach is often preferred for many-body quantum transport because it more effectively handles strong Coulomb correlations in the system. Wang et al. \cite{wang2014nonequilibrium} and Li et al. \cite{li2005quantum} also discuss the regimes in which the NEGF and master equation approaches have been used. They reported that if one needs to systematically improve the weak-coupling approximation and account for strong Coulomb correlations, the master equation approach offers significant advantages. 

To the best of authors' knowledge, no prior studies have investigated thermoelectric transport in QD nanojunctions arranged in an AB ring under weak coupling conditions using the PME with a many-body analysis. Therefore, we are motivated to explore thermoelectric transport, focusing on weakly coupled QDs in an AB ring. We note that, however there are ample of studies on thermoelectric transport in QD based nano-junction in weak coupling regime using PME but without considering AB effect.  For instance, Tagani et al. investigate thermoelectric effects in weakly coupled double QDs using the quantum master equation approach \cite{tagani2012thermoelectric}. Dubi et al. also explore thermospin effects in a QD system connected to ferromagnetic leads and thermoelectric effect in nanoscale junction, employing the master equation approach \cite{dubi2009thermospin,dubi2009thermoelectric}.

In the weak couplig regime, the interaction between the QD system and the electrodes is relatively weak compared to other energy scales, allowing for the use of perturbative methods and simplifying the mathematical treatment. Additionally, understanding the Coulomb blockade effect and sequential tunneling in weak coupling is essential for practical applications in quantum computing and energy-efficient electronics, as it enables precise manipulation of individual electrons. Furthermore, in the weak coupling regime, the transport behavior is primarily determined by the intrinsic properties of the system itself, with minimal influence from changes in the electrodes or their interfaces. This contrasts with the strong coupling regime, where transport is heavily influenced by the interaction between the electrodes, systems, and their interfaces. Hence, we are interested in studying weak coupling regime.

In this article, we investigate the thermoelectric transport properties of triple QDs embedded within an AB ring, connected to two non-ferromagnetic electrodes where the Fermi golden rule is valid for dot-electrode interactions. One QD is coupled to the left electrode, and the third QD is coupled to the right electrode, as shown in Fig. \ref{str_diagrm}. This setup enables simultaneous measurement of electron transport through two independent paths, which enclose a magnetic flux $\Phi$ when a perpendicular field is applied, leading to AB oscillations \cite{bachtold1999aharonov,bruder1996aharonov,yeyati1995aharonov}. Using the PME methods and linear response theory within the Coulomb blockade regime, we calculate the thermoelectric coefficients such as electrical conductance, electronic thermal conductance, thermopower, and figure of merit as functions of chemical potential and magnetic flux at a very low temperature ($1~\text{K}$). In the low-temperature regime, the system accounts for more realistic Coulomb interactions, ensuring that our results are both experimentally relevant and practically applicable. As expected, the electrical conductance oscillates with the magnetic flux, with a period of 2$\pi$ and the magnitude of conductance is suppressed except when the phase values are at $\phi$=(2n+1)$\pi$ (where n=0,1,2,...). Moreover, as the magnetic flux changes from $0$ to $\Phi_0$ ($\Phi_{0}$=$\frac{h}{e}$, one quantum of flux), the electrical conductance experience a 250-fold increase whereas thermoelectric figure of merit experience a 200-fold increase. The effect of onsite and inter-site Coulomb interactions reveals that achieving optimal $ZT$ requires moderate onsite and minimal inter-site Coulomb repulsion. We also briefly discuss how the dot-electrode coupling strength and the asymmetric arrangement of QDs in the ring affect $ZT$. Furthermore, we investigate how increasing the number of QDs in the ring enhances thermoelectric performance.

This paper is structured as follows: In Section II, we present the theoretical model and basic analytical formulae. Section III is dedicated to the presentation of corresponding numerical results and discussions on AB oscillations and thermoelectric transport properties. Finally, in Section IV, we provide a summary of our findings and draw general conclusions based on the results presented in the preceding sections.

\section{Theoretical Model and Formalism}
\subsection{Model}
\begin{figure}
	\centering
	\includegraphics[width=1.0\columnwidth]{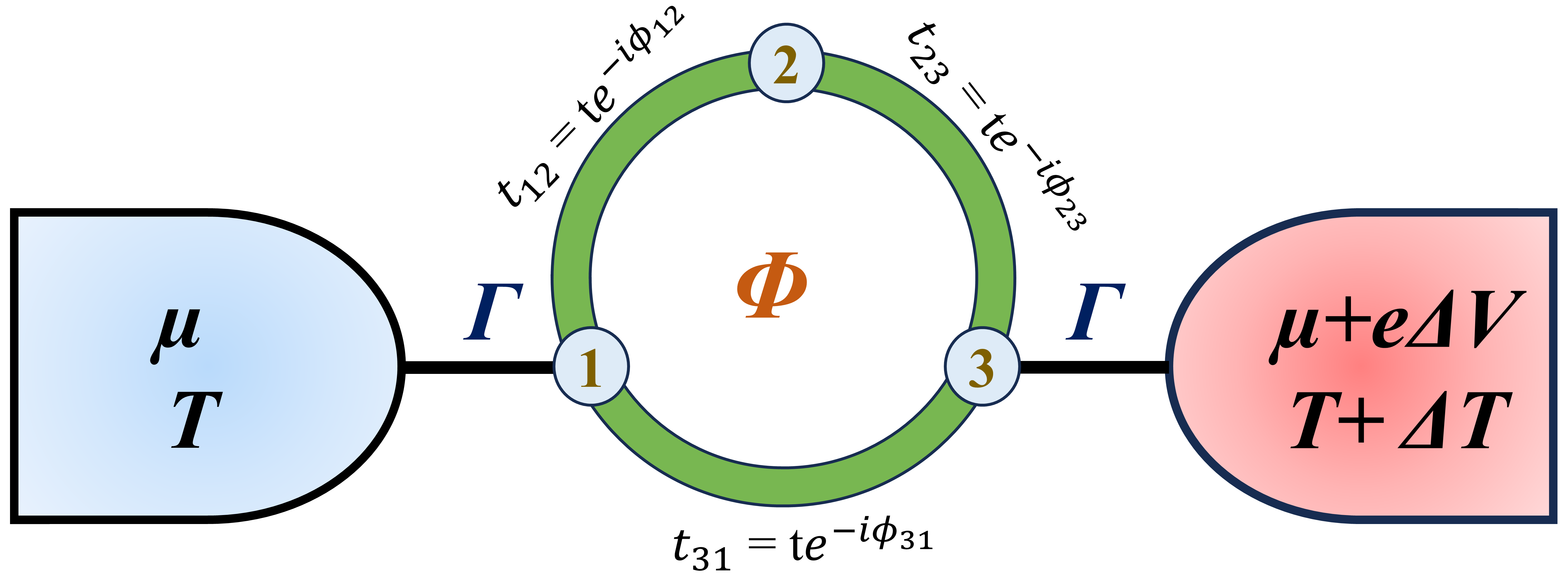}
	\caption{\label{str_diagrm} A schematic of a triple quantum dot Aharonov-Bohm interferometer is shown, where dots 1 and 3 are connected to the electrodes, and a magnetic flux $\Phi$ pierces the AB ring perpendicularly, leading to the induction of quantum interference effects.}
\end{figure}

We consider a triple QDs system  embedded in an AB ring weakly coupled to two non-ferromagnetic electrodes as depicted in Fig. \ref{str_diagrm}. This system can be effectively described by the Hamiltonian $H=H_{lead}+H_D+H_T$. The initial term ($H_{lead}$) in the Hamiltonian represents the noninteracting electrons within the metallic electrodes
\begin{eqnarray}
	H_{lead} = \sum_{\alpha k\sigma}\epsilon_{k\sigma}c^{\dag}_{\alpha k\sigma}c_{\alpha k\sigma},
\end{eqnarray}
where, $\epsilon_{k\sigma}$ denotes the energy of the electron with wave vector $k$, spin $\sigma$ in the electrode, $\alpha$=$L$ (left)/$R$ (right), and  $c^{\dag}_{\alpha k\sigma} (c_{\alpha k\sigma}$) are creation (annihilation) operator of the electron. The second term $H_D$ comprising the interacting extended Hubbard Hamiltonian for isolated 3-membered ring QDs. Taking into consideration both the on-site electron-electron interactions and hopping between nearest neighbor sites, the Hamiltonian can be expressed as follows:
\begin{eqnarray}
	H_D =  \sum_{i=1, \sigma}^N\ \epsilon_{i\sigma} a^{\dag}_{i\sigma}a_{i\sigma} +\sum_{<ij>} - (t_{ij}a^{\dag}_{i\sigma}a_{j\sigma}+h.c.)
	+U\sum_{i=1}^N(n_{i\uparrow}-\frac{1}{2})(n_{i\downarrow}-\frac{1}{2}) \nonumber\\
	+V\sum_{i=1}^N(n_{i\uparrow}+n_{i\downarrow}-1)(n_{i+1\uparrow}+n_{i+1\downarrow}-1),
	\label{exhub}
\end{eqnarray} 

where $\epsilon_{i\sigma}$ is the on-site energy at the discrete site $i$ ($i$=$1$ to $N$, N = no. of sites/QDs) for $\sigma$=$\uparrow$($\downarrow$), and the value of $\epsilon_{i\sigma}$ is set to zero for all sites. $a_{i \sigma}^\dagger$ ($ a_{i \sigma}$) is the creation (annihilation) operator of spin $\sigma$ in the $i^{th}$ site. The notation $<ij>$ signifies that the summation is confined to pairs of nearest neighbor sites. The Hamiltonian ($H_D$) is modeled in a way that exhibits particle-hole symmetry, with the half-filled case becoming the ground state \cite{begemann2008symmetry,darau2009interference}. The hopping integral in the presence of magnetic field $t_{ij}$ is defined in terms of the AB phase ($\phi_{ij}$), denoted as $t_{ij}$ = $te^{-\iota\phi_{ij}}$,  where $t$ represents the nearest-neighbor hopping energy in the absence of magnetic field. The magnetic flux $\Phi$ passing through the ring induces a phase angle $\phi$ which can be written in the following relation \cite{zheng2012thermoelectric}
\begin{eqnarray}
	\phi=\phi_{12}+\phi_{23}+\phi_{31}=\frac{2\pi\Phi}{\Phi_0}.
\end{eqnarray} 
The total magnetic flux enclosed by this circular AB ring is given by $\Phi$ = $\frac{\Phi_{0}\phi}{2\pi}$. The corresponding magnetic field is defined as $B$ = $\frac{\Phi_{0}\phi}{2\pi \Omega}$, where $\Omega$ is the area of the AB ring. For the circle with a radius of $20$ $nm$, the magnetic field ($\frac{\Phi_0}{\Omega}$) is approximately $3.29~T$. Thus, for AB phases of $\phi$ = $\pi/3$  and $\phi$ = $\pi/6$, the corresponding magnetic field values are approximately $0.55~T$ and $0.27~T$, respectively. This range of magnetic field values is also utilized to determine the magnetic field strengths necessary for observing AB oscillations \cite{hod2008magnetoresistance,rai2011magnetic}. $U$ is the Hubbard electron-electron interaction term, $V$ denotes nearest neighbour coulomb repulsion and $n_{i\sigma}$=$a_{i \sigma}^\dagger a_{i \sigma}$. The last term in the Hamiltonian represents the electron tunneling between the electrodes and the QDs. This tunneling process is typically described by a term in the Hamiltonian of the form:
\begin{eqnarray}
	H_T = \Gamma\sum_{\alpha k\sigma}(a^{\dag}_{\alpha\sigma}c_{\alpha k\sigma}+c^{\dag}_{\alpha k\sigma} a_{\alpha\sigma}),
\end{eqnarray}
where $\Gamma$ represent the electrode-dot coupling strengths.
Due to weak coupling to the electrodes, the assumption is made that the potential drop mainly concentrates at the electrode-dot interface, minimizing its impact on the dots themselves. This simplifies analysis, allowing a focus on the electronic behavior within the dots, disregarding detailed electrode potentials.
\subsection{Methods}

\subsubsection{Pauli master equation}
The Liouville equation approach is the preferred method for addressing dynamics in the weak coupling regime. The quantum Liouville equation for the reduced density matrix of the subsystem (in our case it is three QDs) is given by $\frac{\partial \rho}{\partial t}=\frac{i}{\hbar}[\rho,H]+\mathcal{C}\{\rho\}$,
where $\rho$ denotes the reduced density operator, and $\mathcal{C}\{\rho\}$ represents the dissipative component of the dynamics, accounting for the effects after eliminating the electrode memory through the Markovian approximation (temporal coarse-graining)\cite{gebauer2004current}.

Reduced dynamics are often approximated using a generalized quantum master equation, which includes both diagonal and off-diagonal elements of the density matrix to capture coherences between different charge states. However, in the weak coupling regime, the PME is limited to the diagonal elements of the density matrix \cite{fischetti1998theory,fischetti1999master}. These diagonal elements represent the occupation probabilities of the eigenstates of the system Hamiltonian, while off-diagonal terms, which account for coherences between different charge states, are neglected. The PME governs the time evolution of the occupation probabilities, corresponding to the diagonal elements of the reduced density matrix \cite{dubi2009thermospin,hettler2003current}. The conventional method involves solving for the steady-state solution, which reduces the problem to an algebraic system of linear equations. This approach yields the kernel of the rate matrix. In non-stationary scenarios, applying the PME can result in violations of current continuity \cite{gebauer2004current,frensley1990boundary}. Additionally, the use of the PME is generally justified only for very small devices \cite{gebauer2004kinetic,gebauer2004current}. Given the mesoscopic scale of our system, employing the PME is both appropriate and valid in the weak-coupling limit. The PME also can be understood as an outcome of first-order time-dependent perturbation theory (Fermi's golden rule) or as a solution derived from a many-body Schrödinger equation \cite{bruus2004many}.

We use the exact diagonalization method \cite{hasegawa1989hole,bonvca1989exact} to determine the eigenstates $|s\rangle$ and corresponding eigenenergies $E_s$ of the Hamiltonian $H_D$ (with $4^{3}$=$64$ many-body basis states) for triple QDs ring system shown in Fig. \ref{str_diagrm}. The full eigen space (64 eigen vectors and eigen values) of the QDs system is considered for calculating all transport properties within PME formalism. The electrodes are treated as electronic reservoirs, each described by Fermi distributions.
The PME (taking diagonal element of the reduced density matrix ($\rho_{ss}=P_s$)) is written as follows\cite{schaller2014open}:
\begin{eqnarray}
	\frac{dP_s}{dt} = \sum_{s'} \left( W_{s' \rightarrow s} P_{s'} - W_{s \rightarrow s'} P_s \right),
\end{eqnarray}
where $W_{s^\prime \rightarrow s}$ ($W_{s \rightarrow s^\prime}$) denotes the transition rate from the many-body Fock state $|s^\prime\rangle$ to $|s\rangle$ ($|s\rangle$ to $|s^\prime\rangle$), where the states differ by one electron, and $P_s$ ($P_{s^\prime}$) is the probability that the system will be in the many-body state $|s\rangle$ ($|s^\prime\rangle$), and $E_s$ ($E_{s^\prime}$) is the respective energy.
The transition rates are determined as follows:
\begin{eqnarray}
	W_{{s^\prime}\rightarrow{s}}^{L+}=\frac{\Gamma}{\hbar} f_L((E_s-E_{s^\prime})-\mu) \sum_{\sigma} 
	|<s|a^\dag_{1\sigma}|s^\prime>|^2,
\end{eqnarray}
\begin{eqnarray}
	W_{{s^\prime}\rightarrow{s}}^{R+}=\frac{\Gamma}{\hbar} f_R((E_s-E_{s^\prime})-\mu) \sum_{\sigma} 
	|<s|a^\dag_{N\sigma}|s^\prime>|^2.
\end{eqnarray} 
The equations for $W_{{s} \rightarrow {s^\prime}}^{L-}$ and $W_{{s} \rightarrow {s^\prime}}^{R-}$ are obtained by replacing $f_{L,R}((E_s-E_{s^\prime})-\mu)$ with $(1-f_{L,R}((E_s-E_{s^\prime})-\mu))$, where $f_{L/R}$ is the Fermi function for the left/right electrode. In this context, the symbols $+/-$ denotes the creation/destruction of an electron within the dots due to electron migration from/to the $L$/$R$ electrodes. Furthermore, we assume that creation and annihilation occur solely at the sites directly connected to the electrodes. The total transition rate is calculated by summing four terms ($W_{{s} \rightarrow {s^\prime}}=W_{{s} \rightarrow {s^\prime}}^{L+}+
W_{{s} \rightarrow {s^\prime}}^{R+}+W_{{s} \rightarrow {s^\prime}}^{L-}+
W_{{s} \rightarrow {s^\prime}}^{R-}$). The population of many-body states can be determined by solving the steady-state PME, $\frac{dP_s}{dt}= 0$. This equation takes the form of a homogeneous linear system ($AX=0$), which cannot be directly solved. To address this, we utilize the constraint $\mathrm{\sum_s{P_s}=1}$ to eliminate one row/column of the matrix. This allows us to reformulate the eigenvector problem into an inhomogeneous linear system ($AX=B$), which can be solved using linear algebraic methods.
\subsubsection{Charge and heat current}
In a steady state, the current flowing from the QD to one electrode is precisely balanced by the current flowing from the other electrode into the QD, preventing any net charge accumulation in the QD. However, when considering the current between the QD and a single electrode, this current does not vanish. This is because it represents the actual flow of electrons through the system. The charge and heat current through the system can thus be expressed as:
\begin{eqnarray}
	I_{\alpha} =e \sum_{s,s^\prime}(W_{{s^\prime} \rightarrow s}^{\alpha+}
	P_{s^\prime}-W_{{s} \rightarrow {s^\prime}}^{\alpha-} P_s),
\end{eqnarray}
\begin{eqnarray}
	Q_{\alpha} = \sum_{s,s^\prime}((E_s-E_{s^{'}})-\mu)(W_{{s^\prime} \rightarrow s}^{\alpha+}
	P_{s^\prime}-W_{{s} \rightarrow {s^\prime}}^{\alpha-} P_s),
\end{eqnarray}
where $\mu$ is the chemical potential of the electrodes.
\subsubsection{Thermoelectric formulation}
In thermoelectric investigations, especially within the linear response regime, we consider small deviations from equilibrium. This involves assuming a small temperature difference between the electrodes and a small voltage difference across them. Specifically, if the left electrode is slightly hotter than the right one, i.e., $T_L=T_R+\Delta T$ and $\mu_L=\mu_R-e\Delta V$, then the expressions for charge and heat currents can be derived using linear response theory written as follows:
\begin{eqnarray}
	I_{\alpha} =G_V \Delta V+G_T \Delta T,
\end{eqnarray}
\begin{eqnarray}
	Q_{\alpha} =M \Delta V+K \Delta T,
\end{eqnarray}
where $G_V$ represents the electrical conductance and $G_T$ refers to the thermoelectric coefficient, which is associated with the charge current response to the heat flux. We obtain $G_V$ and $G_T$ by setting $I_L=\frac{1}{2}(I_L-I_R)$ and expanding the Fermi-Dirac distribution function as $f_L(x) = f_R(x) - \frac{(x-\mu)}{T} f^{'}(x) \Delta T+ e \Delta Vf^{'}(x)$. where $f^\prime  (x)=\frac{\partial f(x)}{\partial x}$ .
\begin{eqnarray}
	G_V =\frac{e^{2}\Gamma}{2\hbar}\sum_{s,s^\prime}\sum_{\sigma} f^\prime  ((E_s-E_{s^\prime})-\mu)|<s|a^\dag_{1\sigma}|s^\prime>|^2  (P_s+P_{s^\prime}),
	\label{gv}
\end{eqnarray}
\begin{eqnarray}
	G_T =\frac{-e\Gamma}{2\hbar}\sum_{s,s^\prime}\sum_{\sigma} f^\prime  ((E_s-E_{s^\prime})-\mu)\frac{((E_s-E_{s^\prime})-\mu)}{T} 
	|<s|a^\dag_{1\sigma}|s^\prime>|^2 (P_s+P_{s^\prime}).
	\label{gt}
\end{eqnarray} 
Under the condition of zero charge current ($I_\alpha=0$), the thermopower can be calculated as $S=\frac{G_T}{G_V}$. Using the Onsager relation, the electron thermal conductance can be written as: \cite{zianni2008theory},
\begin{eqnarray}
	k_e=\frac{-Q_\alpha}{\Delta T}|_{I_ \alpha=0}\nonumber \\
	=K-S^{2}G_{V}T,
\end{eqnarray}
where,
\begin{eqnarray}
	K =\frac{\Gamma}{2\hbar}\sum_{s,s^\prime}\sum_{\sigma} f^\prime  ((E_s-E_{s^\prime})-\mu) \frac{((E_s-E_{s^\prime})-\mu)^2}{T}
	|<s|a^\dag_{1\sigma}|s^\prime>|^2 (P_s+P_{s^\prime}).
\end{eqnarray}
The figure of merit can be expressed as $ZT=\frac{S^{2}G_{V}T}{k_e+k_{ph}}$, where $k_{ph}=3k_0$ and $k_0=(\pi^2 k_B^2 T)/3h$ represents the quantum thermal conductance which is universal and gives a reasonable value for nanoscale junctions\cite{schwab2000measurement,dubi2009thermospin}.
\section{Results and discussions}

\subsection{Three QDs arranged symmetrically in an AB ring}
This section focuses on the symmetric arrangement of triple QDs in an AB ring. We adopt equal AB phase for each arm, which can be expressed as follows: $\phi_{12}=\phi_{23}=\phi_{31}=\frac{\phi}{3}=\frac{2\pi\Phi}{3\Phi_0}$. This AB phases can be obtained by considering a vector potential of the type \(\vec{\mathbf{A}} = \left(-\frac{By}{2}, \frac{Bx}{2}, 0\right)\), as briefly discussed in section I of the supplementary information.
Following the aforementioned theory, in the subsequent calculation, we utilize a nearest-neighbor hopping energy ($t$) of $1~\text{meV}$. Additionally, we assume symmetric coupling between the QDs and the two electrodes  and take  $\Gamma (0.001~\text{meV}) \ll k_{B}T$ which signifies the weak coupling regime, where the system interacts with the electrodes so weakly that it doesn't significantly affect the overall temperature of the environment. This indicates that the energy exchange between the system and the electrodes due to thermal fluctuations is relatively small compared to the thermal energy of the environment.

\subsubsection{Aharonov-Bohm oscillations}  

\begin{figure}
	\centering
	\includegraphics[width=0.8\columnwidth]{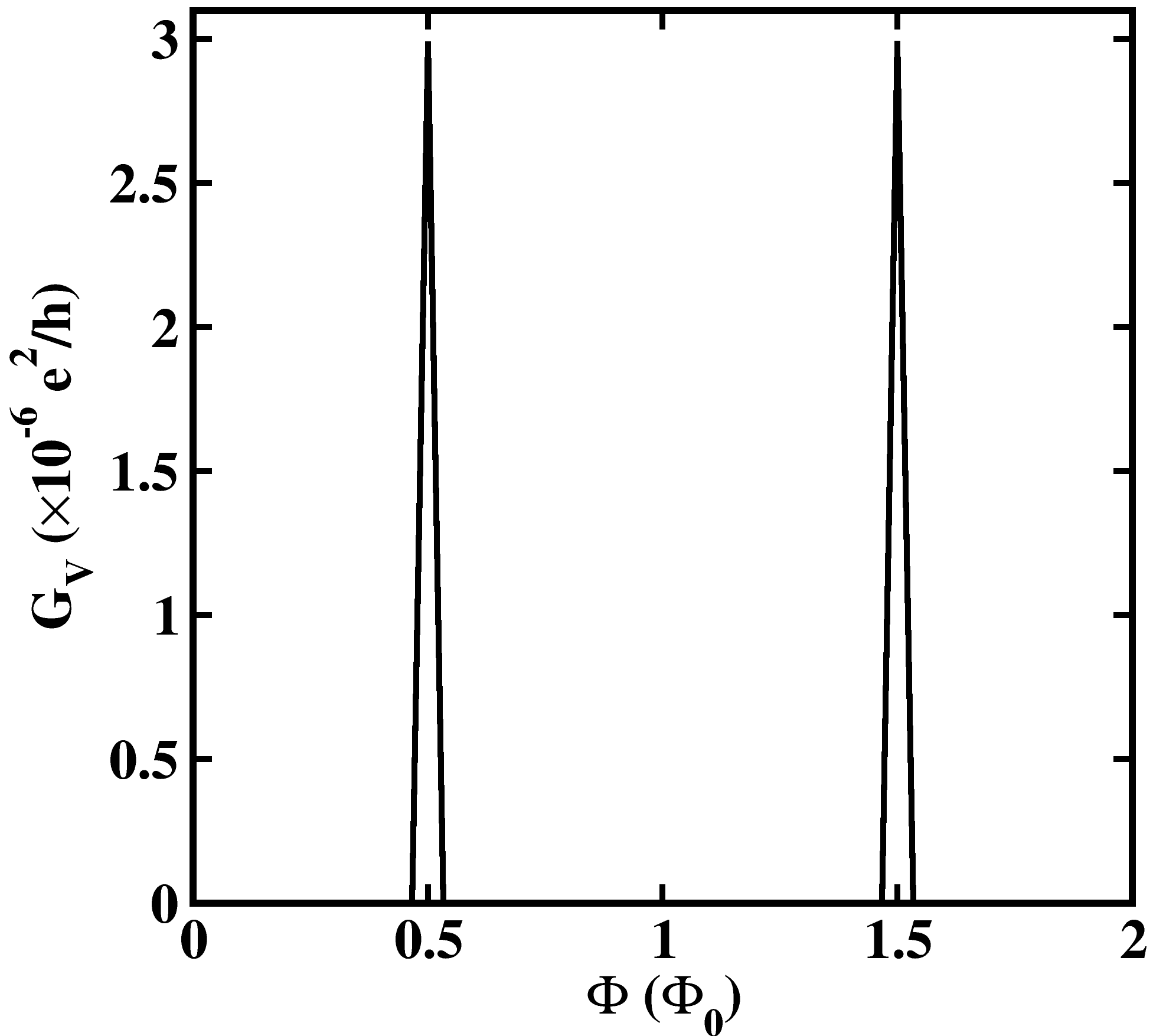}
	\caption{\label{gv_phi}Electrical conductance as a function of $\Phi$ showing periodic Aharonov-Bohm oscillation at $\mu=0.0~\text{meV}$, $\Gamma=0.001~\text{meV}$, $k_{B}T=0.086~\text{meV}$ (T=$1~\text{K}$), $t=1~\text{meV}$, $U=5~\text{meV}$ and $V=1~\text{meV}$.}
\end{figure}

At first, we study the well established AB oscillation before studying thermoelectric properties. In the three QDs ring structure, as expected the electrical conductance oscillates periodically with $\Phi$ known as AB oscillation, is shown in Fig. \ref{gv_phi}. Each QD serves as a site where electrons can localize, forming discrete energy levels. As the magnetic flux varies, the relative phases of the electron waves interfere constructively or destructively, leading to oscillations in the conductance.

In our case, with the absence of magnetic flux ($\Phi = 0$) for $\mu = 0$, 3-electron ($3e^{-}$) state, $|\Psi_{3e^-}^{gs}\rangle$  is the ground state and it's probability ($P_3$) is $1$. When the chemical potential ($\mu$) aligns with the transition energy, resonant tunneling occurs. This alignment facilitates more efficient electron tunneling through the nanojunction, resulting in enhanced conductance and the formation of sharp peaks in $G_V$.  At $\Phi = 0$, the ground state energy for the  $3e^{-}$ state is $E_3=-4.87~\text{meV}$, while for the $2e^{-}$ state, it is $E_2=-2.65~\text{meV}$ (see the Table S I in the supplementary information for important energy levels). The transition energy between $3e^{-}$ and  $2e^{-}$ ground state ($E_3-E_2$) is $-2.22~\text{meV}$, and similarly, the corresponding transition energy between $3e^{-}$ and $4e^{-}$ state is $2.22~\text{meV}$. These values are significantly away from $\mu=0$ suggests that transitions between the $2e^{-}$ and $3e^{-}$ states or between the $3e^{-}$ and $4e^{-}$ states are not energetically favorable. Thus, antiresonance occurs, leading to the absence of a peak in $G_V$.

However, when the magnetic flux through the system reaches certain values, such as $\Phi$ = $\frac{\Phi_{0}}{2}$, the $G_V$ exhibits peaks (although small values), indicating AB oscillations. This phenomenon occurs due to the quantization of energy levels of electrons when a perpendicular magnetic field is applied. As the magnetic field strength increases, the Landau levels of electrons become more quantized, leading to oscillations in conductance as a function of $\Phi$. For $\Phi$ = $\Phi_{0}/2$, the gap between transition energy and chemical potential ($\mu=0$) is least and hence $G_V$ shows peaks but very small values. The modification of the energy levels has a significant impact on the transmission probabilities between the $2e^{-}$ and $3e^{-}$ states. By tuning the energy gap between these states, the magnetic flux influences the probabilities of electron transitions between them. As the magnetic flux increases from $\Phi = 0$ to $\Phi$ = $\Phi_{0}/2$, we observe a slight decrease in $P_3$ ($P_3$=$0.98$), accompanied by small contributions appearing in $P_2$ ($P_2$=$0.02$). This alteration brings the transition energy $E_3(-5.05~\text{meV})-E_2(-3.95~\text{meV})=-1.10~\text{meV}$ closer to $\mu=0$, thereby  the transition between the $3e^{-}$ and the $2e^{-}$ state occurs. Consequently, a weak interference between the $2e^{-}$ and $3e^{-}$ states occurs as the probability $P_2$ is much less compared to $P_3$, resulting in a small conductance value of $\approx$ $3\times10^{-6}~e^{2}/h$.

Moreover, the derivative of the Fermi-Dirac distribution ($f^{\prime}((E_{s}-E_{s{\prime}})-\mu)$) plays a crucial role in determining the conductance of a system. Thus, this minute conductance also arises from the tail part of the $f^{\prime}((E_{3}-E_{2})-\mu)$, at low temperature of the system as the $G_V$ is directly proportional to the $f^{\prime}((E_{3}-E_{2})-\mu)$ (as per Eq. \ref{gv}). Notably, at very lower temperatures, the conductance value decreases further, as the derivative of the Fermi-Dirac distribution tends toward a perfect Dirac delta function.

At $\Phi$ =$\Phi_{0}$ ($\phi$=$2\pi$), conductance oscillations repeat due to interference of electron wavefunctions encircling QDs in the magnetic field. Following this repetition, AB oscillations again manifest as peaks in the $G_V$ at $\Phi$ = $\frac{3}{2}\Phi_{0}$, indicating another cycle of constructive interference and emphasizing the periodic nature of the phenomenon. The magnitude of conductance is suppressed except when the phase values are at $\phi$=(2n+1)$\pi$ (where n=0,1,2,...). These AB oscillations provide insight into the quantum mechanical behavior of electrons in mesoscopic systems and are crucial for understanding and engineering nanoscale electronic devices.

\subsubsection{Thermoelectric transport}
\begin{figure}
	\centering
	\includegraphics[width=1.0\columnwidth]{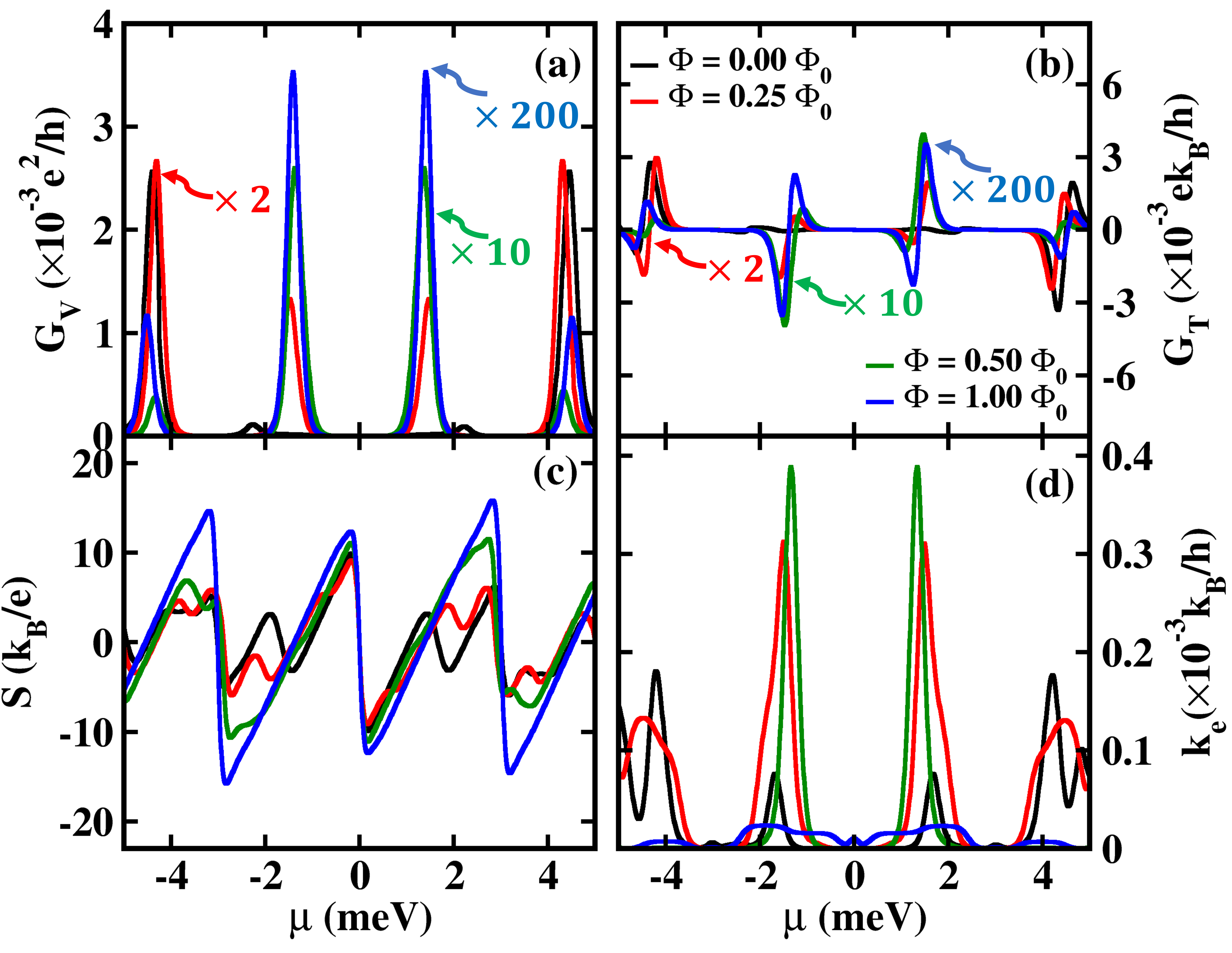}
	\caption{\label{all_thermo}  (a)Electrical conductance ($G_V$), (b)thermoelectric coefficient ($G_T$), (c)Seebeck coefficients ($S$), and electronic thermal conductivity ($k_e$) respectively, as functions of $\mu$ for different magnetic fluxes ($\Phi$) with $\Gamma=0.001~\text{meV}$, $k_{B}T=0.086~\text{meV}$ (T=$1~\text{K}$), $U=5~\text{meV}$ and $V=1~\text{meV}$ for triple QDs embedded in an AB ring. The plot for $\Phi$=0.25$\Phi_0$, $\Phi$=0.50$\Phi_0$ and $\Phi$=$\Phi_0$ (red, green and blue line) has been scaled to multiples of 2, 10 and 200 in $G_V$ and $G_T$  values respectively.}
\end{figure}

For thermoelectric transport study, we analyze the behavior of electrical conductance ($G_V$), thermoelectric coefficient ($G_T$), thermopower ($S$), electronic thermal conductance ($k_e$), and the figure of merit ($ZT$) as a functions of the chemical potential ($\mu$) and magnetic flux ($\Phi$). We here set $U=5~\text{meV}$ and $V=1~\text{meV}$. In the above section, we have already explained the AB oscillation in $G_V$ at $\mu=0$ and found very small conductance value ($\approx$$3\times10^{-6}~e^{2}/h$). Here $G_V$ can be tuned by varying $\mu$ at different $\Phi$ which is given in Fig. \ref{all_thermo} (a). Here, the presence of four resonance peaks in $G_V$ indicates a complex interplay of quantum mechanical effects within the system.

In the absence of magnetic flux, the first peak is observed when \(\mu \approx -4.45~\text{meV}\), which corresponds to the transition energy between the \(1e^{-}\) and \(2e^{-}\) states. At this point, the occupation probabilities of the \(1e^{-}\) and \(2e^{-}\) states are equally probable. As \(\mu\) increases, a second peak appears at \(\mu \approx -2.22~\text{meV}\). This aligns with the transition energy between the \(2e^{-}\) and \(3e^{-}\) states (\(E_3 - E_2 = -2.22~\text{meV}\)), where the probabilities of the \(2e^{-}\) state (\(P_2\)) and the \(3e^{-}\) state (\(P_3\)) are exactly equal, i.e., \(P_2 = P_3 = 0.5\). In this scenario, resonance occurs. According to the detailed balance condition, in the thermal equilibrium, the rates of forward and reverse processes are equal, the probabilities of transitions between these states are equal. Therefore, \(P_2\) and \(P_3\) are equally probable, leading to peaks in conductance. The derivative of the Fermi function, \(f'((E_s - E_{s'}) - \mu)\), shows a peak when the chemical potential equals the transition energy. Since \(G_V\) depends on \(f'((E_s - E_{s'}) - \mu)\), it shows a peak at resonant tunneling \((E_s - E_{s'} = \mu)\). Similarly, the third and fourth peaks occur when \(\mu = 2.22~\text{meV}\) and \(\mu = 4.45~\text{meV}\), corresponding to the transitions from the \(3e^{-}\) to \(4e^{-}\) and \(4e^{-}\) to \(5e^{-}\) states, respectively. Thus, $G_V$ can be enhanced by meeting the resonant condition, which is achieved through tuning of $\mu$.
For instance, the value of $G_V$ is increased significantly to  2.68 $\times$ 10$^{-3}$~e$^{2}/h$ at resonant ($\mu$ = -4.45 meV) compared to  1.49 $\times$ 10$^{-9}$~e$^{2}/h$ at $\mu$=0. Moreover, there is further scope of enhancing this value with finite magnetic flux which is described below.

When the magnetic flux \(\Phi\) is equal to \(\Phi_{0}\), the \(G_V\) exhibits four symmetric peaks at \(\mu \approx -4.51, -1.40, 1.40,\) and \(4.51~\text{meV}\). These peaks correspond to the transitions between the \(1e^{-}\) and \(2e^{-}\) states, the \(2e^{-}\) and \(3e^{-}\) states, the \(3e^{-}\) and \(4e^{-}\) states, and the \(4e^{-}\) and \(5e^{-}\) states, respectively, where these transitions are energetically favorable. Moreover, Fig. \ref{all_thermo}(a) reveals a substantial enhancement in $G_V$ with increasing magnetic flux $\Phi$. $G_V$ can reach upto $0.705~e^{2}/h$ at $\Phi = \Phi_{0}$ compared to 2.68 $\times$ 10$^{-3}$~e$^{2}/h$ at $\Phi$ = 0. The overall $G_V$ experiences approximately 250-fold increase when $\Phi$ changes from $0$ to $\Phi_{0}$. This remarkable increase in $G_V$ with increasing $\Phi$ can be attributed to the enhancement of transition matrix elements at resonance. The enhancement of transition matrix elements leading to strong interference effect. The details of important eigenstates of the electronic system and the transition matrix elements between them for both $\Phi=0$ and $\Phi=\Phi_{0}$ are given in section III and section IV of the supplementary information, respectively.

At resonant values of $\mu$, $P_s$ = $P_{s^\prime}$ = 0.5 (see the Fig. S2 in supplementary information) and  $f^{\prime}((E_{s}-E_{s^{\prime}})-\mu) = \frac{-1}{4k_{B}T}$ = $-2.90~(\text{meV})^{-1}$ at a low temperature of $1~\text{K}$. Thus, $G_V$ in equation \ref{gv} can be simplified as $G_V$=$-\pi\times e^{2}/h\times\Gamma\times -2.90 \times \sum_{\sigma}|\langle s|a^\dag_{1\sigma}|s^\prime\rangle|^{2}\times(0.5+0.5)$.   For \(\Phi = 0\), the corresponding squared transition matrix elements are \(\sum_{\sigma} \left|\langle \Psi_{2e^{-}}^{1es} | a_{1\sigma}^{\dagger} | \Psi_{1e^{-}}^{2es} \rangle \right|^{2}\)=0.201 and \(\sum_{\sigma} \left|\langle \Psi_{3e^{-}}^{gs} | a_{1\sigma}^{\dagger} | \Psi_{2e^{-}}^{1es} \rangle \right|^{2}\) = 0.012, at \(\mu = -4.45~\text{meV}\) and \(\mu = -2.22~\text{meV}\), yielding conductance values of approximately \(2.68 \times 10^{-3}~e^{2}/h\) and \(1.12 \times 10^{-3}~e^{2}/h\), respectively. Similar things also occurs  at \(\mu = 2.22~\text{meV}\) and \(\mu = 4.45~\text{meV}\) in the electron doping regime. In contrast, for \(\Phi = \Phi_{0}\), the scenario changes. For hole doping (\(\mu = -4.51~\text{meV}\) and \(\mu = -1.40~\text{meV}\)), the squared transition matrix elements, \(\sum_{\sigma} \left|\langle \Psi_{2e^{-}}^{1es} | a_{1\sigma}^{\dagger} | \Psi_{1e^{-}}^{2es} \rangle \right|^{2}\) and \(\sum_{\sigma} \left|\langle \Psi_{3e^{-}}^{gs} | a_{1\sigma}^{\dagger} | \Psi_{2e^{-}}^{1es} \rangle \right|^{2}\), significantly increase to 25.57 and 77.38, respectively, leading to conductance values of \(0.233~e^{2}/h\) and \(0.705~e^{2}/h\). The same trend is observed in the electron doping regime (\(\mu = 1.40~\text{meV}\) and \(\mu = 4.51~\text{meV}\)). A detailed analysis of these transition matrix elements is provided in case-2 of section IV in the supplementary information.

Indeed, the observation of 250-fold increased conductance values with the tuning of perpendicular magnetic flux from $0$ to $\Phi_{0}$\ is noteworthy. When $\Phi=0$, the transition matrix elements  are very low, resulting in weak interference. In contrast, for $\Phi=\Phi_0$, the transition matrix elements are relatively high compared to $\Phi=0$, leading to strong constructive interference between states that actively participate in transport and enhancing the peak in $G_V$. Conversely, at other $\mu$ values where antiresonance occurs, known as destructive interference, the conductance is diminished. Therefore, the observed resonance peaks in the conductance spectra arise from both resonant tunneling through the discrete energy levels of the QDs and the tuning of perpendicular magnetic fluxes. 
Several studies have also demonstrated that increasing the AB flux in QD junctions can enhance thermoelectric coefficients. For instance, Kim et al. found that, electrical conductance can increases by one order of magnitude by applying AB flux  in an AB interferometer with an embedded QD in the Kondo regimeat low temperatures \cite{kim2003thermoelectric}. Similarly, Zheng et al. reported that the conductance also rises with increasing flux in an AB ring embedded with a QD \cite{zheng2012thermoelectric}.  

Additionally, the electrical conductance in Fig. 3(a) exhibits symmetrical behavior at the middle of the Coulomb gap ($\mu$=0) due to the particle-hole symmetry in our Hamiltonian. Similarly, Begemann et al. (Fig. 2 of Ref. \cite{begemann2008symmetry}) also reported that differential conductance remains symmetric about zero gate voltage due to the presence of particle-hole symmetry. However, Fig. 5 of Chen et al.\cite{chen2015quantum} shows that when the middle of the Coulomb gap is used as a reference point, the electrical conductance spectrum lacks mirror symmetry. This asymmetry may result from the breaking of particle-hole symmetry, potentially due to variations in onsite energy or additional interactions that deviate from strict particle-hole symmetry. 

Another observation from the electrical conductance plots (Fig. S5 in the supplementary material) for different onsite energy values for $\Phi$=$\Phi_{0}$ is that as the onsite energy increases, the peak positions shift while the peak heights remain unchanged. This shift results from changes in onsite energy, which modify the energy levels, thus altering the resonance conditions. However, since the transition matrix elements governing electron transport remain constant, the peak heights in the conductance spectrum are unaffected. Additionally, Fig. S6(a) in the supplementary material shows that as temperature increases, the conductance resonance peaks diminish and broaden due to the influence of the Fermi–Dirac distribution. At low temperatures, this distribution sharply defines the occupancy of energy levels up to the chemical potential. As the temperature rises, the energy distribution widens, allowing electrons to occupy a broader range of states. Consequently, the initially sharp conductance resonance peaks become less distinct and spread over a wider energy range. This broadening also leads to a reduction in the amplitude of the AB oscillations, as thermal fluctuations allow multiple electronic states to contribute to the transport process.

In Fig. \ref{all_thermo} (b), the behavior of the thermoelectric coefficient ($G_T$) is depicted as a function of $\mu$ at varying magnetic fluxes. At the point where $G_V$ peaks, both $G_T$ and $S$ gives zero, signifying an electron transition at that specific energy level, leading to an electrical current but no net energy transport. This occurrence arises when the probability of occupying the $N$ and $N + 1$ electron states is equal ($P_N$=$P_{N+1}$), aligning their transition energy precisely with $\mu$ of the electrode. In the above discussed scenario, we have already elucidated the contribution of electronic states to the transport process. Mathematically, at resonance, the value of $G_T$ vanishes, as indicated by equation \ref{gt}. Notably, the temperature gradient fails to produce a significant charge current at resonance energies.

From equation \ref{gt}, we observe that $G_T$ exhibits peaks under off-resonant conditions with an optimal gap between the transition energy ($E_{s}-E_{s^{\prime}}$) and the chemical potential ($\mu$). The value of $G_T$ depends on $f'((E_s - E_{s'}) - \mu)$ and the difference $(E_s - E_{s'}) - \mu$ apart from square of transition matrix element. When $f'((E_s - E_{s'}) - \mu)$ is at its maximum, which happens at resonance (i.e., $E_s - E_{s'} = \mu$), $G_T$ becomes zero. On the other hand, if the gap between $E_s - E_{s'}$ and $\mu$ is too large (strong off-resonant condition), $f'((E_s - E_{s'}) - \mu)$ approaches zero. Therefore, a moderate gap between the transition energy and $\mu$ is needed to achieve a peak in $G_T$. For $\Phi = 0$ and $\mu = -4.35~\text{meV}$, $G_T$ reaches a peak of $0.0028~ek_{B}/h$. In contrast, for $\Phi = \Phi_0$ and $\mu = -4.38~\text{meV}$, $G_T$ sharply rises to $0.17~ek_{B}/h$. In both cases, the transport involves only the \(1e^{-}\) and \(2e^{-}\) states. This significant increase in $G_T$ is driven by the modulation of energy eigenstates due to the magnetic flux, which increases the transition matrix elements from 0.201 to 25.57 (see case-2 of section IV in the supplementary information for transition matrix elements). Similar behavior is observed for other peaks in $G_T$. The observed increase in $G_T$ with increasing $\Phi$, underscores the influence of interference patterns modulated by the AB phase.

Moving to Fig. \ref{all_thermo} (c), the thermopower ($S$) as a function of $\mu$ and $\Phi$ exhibits numerous zero points, indicating regions where the voltage generated by a temperature gradient across the system is zero. These zero points correspond to specific features in the conductance spectrum, including all resonant points and dip points. Furthermore, as the electrode temperature varies, the thermal energy fluctuates, resulting in a broadening or narrowing of the energy levels in the QD nanojunction. Consequently, the energy window over which electrons can contribute to the Seebeck effect widens or narrows, thereby decreasing or increasing the Seebeck coefficient. A broader energy range reduces the effectiveness of energy filtering, leading to a decrease in the magnitude of the Seebeck coefficient, while a narrower energy range enhances the Seebeck effect, beneficial for thermoelectric transport. Additionally, the sign of $S$ provides information about the dominant charge carriers in the system. A positive $S$ indicates that the majority of carriers are holes, implying that the flow of positive charge carriers contributes more to the thermoelectric response. Conversely, a negative $S$ suggests that electrons are the dominant carriers, with negative charges contributing more significantly to the thermoelectric behavior. Furthermore, an increase in $\Phi$ leads to an increase in the absolute value of $S$.
\par
Figure \ref{all_thermo}(d) depicts the electronic thermal conductance ($k_e$) as a function of $\mu$ at different $\Phi$. Indeed, with increasing temperatures, the $k_e$ tends to increase. However, to achieve good thermoelectric performance in a QD junction, it is essential to reduce $k_e$ while maximizing $G_V$ and  $S$. This strategy is particularly effective when operating at lower temperatures, as it allows for a more favorable temperature gradient across the device, resulting in more efficient conversion of heat into electrical energy.

Operating at lower temperatures is beneficial for several reasons. Firstly, it helps to maintain the integrity of the QD nanojunction. Higher temperatures can lead to increased electron-phonon interactions, which can degrade the performance of devices over time. Secondly, at lower temperatures, the thermoelectric properties of the system can be more precisely controlled and optimized.

In our observations, $k_e$ at $1~\text{K}$ is advantageous for adjusting the thermoelectric properties of the triple QDs embedded in an AB ring, providing both theoretical and experimental validation. Additionally, fluctuations in $k_e$ increase with $\Phi$ except for $\Phi = \Phi_{0}$, indicating greater variability in heat transport properties as the AB phase changes. The very low value of $k_e$ observed at $\Phi = \Phi_{0}$ suggests that thermoelectric performance should increase under these conditions. These fluctuations may offer opportunities for controlling and optimizing the thermoelectric behavior of the system.

\par

\begin{figure}
	\centering
	\includegraphics[width=0.8\columnwidth]{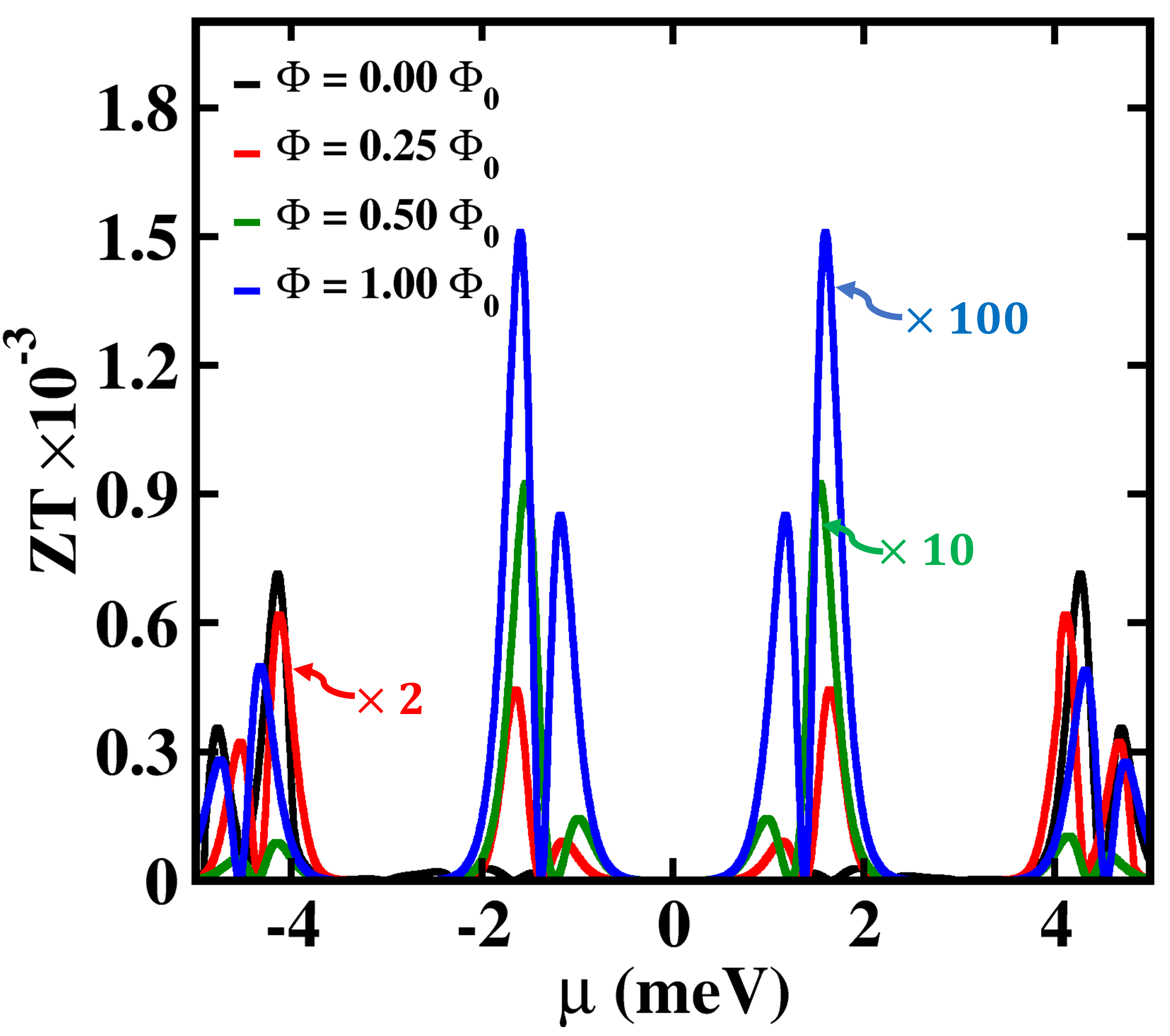}
	\caption{\label{zt_all} The thermoelectric figure of merit ($ZT$) as a functions of $\mu$ at different magnetic fluxes ($\Phi$) with $\Gamma=0.001~\text{meV}$, $k_{B}T=0.086~\text{meV}$, $t=1~\text{meV}$, $U=5~\text{meV}$ and $V=1~\text{meV}$. The plot for $\Phi$=0.25$\Phi_0$, $\Phi$=0.50$\Phi_0$ and $\Phi$=$\Phi_0$ (red, green and blue line) has been scaled to multiples of 2, 10 and 100 in $ZT$ values respectively.}
\end{figure}
Finally, we investigate the thermoelectric figure of merit in our system. As shown in Fig. \ref{zt_all}, the figure of merit is zero at resonances and in the Coulomb blockade region, as $ZT$ depends directly on $S$. Due to the inverse relationship between conductance and thermopower, the figure of merit exhibits fluctuating behavior. From the Fig. \ref{zt_all}, we observed that with no flux present in the QD ring, the maximum $ZT$ value is approximately $0.71\times10^{-3}$. However, as we introduce magnetic flux through the QD ring, the $ZT$ value remarkably increases upto two order of magnitude, with a maximum value of $0.15$ when $\Phi$ = $\Phi_0$. Liu et al. \cite{liu2011role} and Zheng et al. \cite{zheng2012thermoelectric} also observed similar trends, finding that the $ZT$ increases with variations in the magnetic flux within the QD ring system. Additionally, Lu et al. reported an increase in $ZT$ by up to an order of magnitude when the magnetic flux was changed from 0 to 0.5$\pi$ \cite{lu2014thermoelectric}. Kuo et al. studied the impact of level degeneracy on the thermoelectric properties of QDs in a nanowire junction under the Coulomb blockade regime. They found that ZT is significantly enhanced with level degeneracy, even in the presence of finite electron hopping in coupled QD systems \cite{kuo2017large}. Chen et al. analyzed quantum interference and orbital filling effects in QD molecules, showing that destructive interference blocks charge transport more than heat transport, with maximum power factor achieved under full-filling in triangular QD molecule and low-filling in serially coupled QDs \cite{chen2015quantum}.

Furthermore, $ZT$ decreases with increasing temperature, as shown in Fig. S6 in the supplementary material. The broadening of conductance peaks reduces electrical conductance, further contributing to the decrease in $ZT$. Additionally, at higher temperatures, increased phonon contributions to thermal conductivity raise total thermal conductance, further lowering $ZT$. Therefore, the low-temperature regime is more favorable for studying transport in QD nanojunctions, where the effects of Coulomb correlation remain significant and relevant for state-of-the-art experimental setups \cite{kouwenhoven1997electron,svilans2016experiments}. The numerical results demonstrate that the symmetrically arranged triple QDs in an AB ring nanojunction exhibits a good thermoelectric figure of merit, suggesting its potential for efficient energy conversion. In the studied system, the $ZT$ values can be attained across a broad range of adjustable parameters. 

\subsubsection{Effect of onsite and inter-site Coulomb interactions on ZT }
\begin{figure}
	\centering
	\includegraphics[width=0.8\columnwidth]{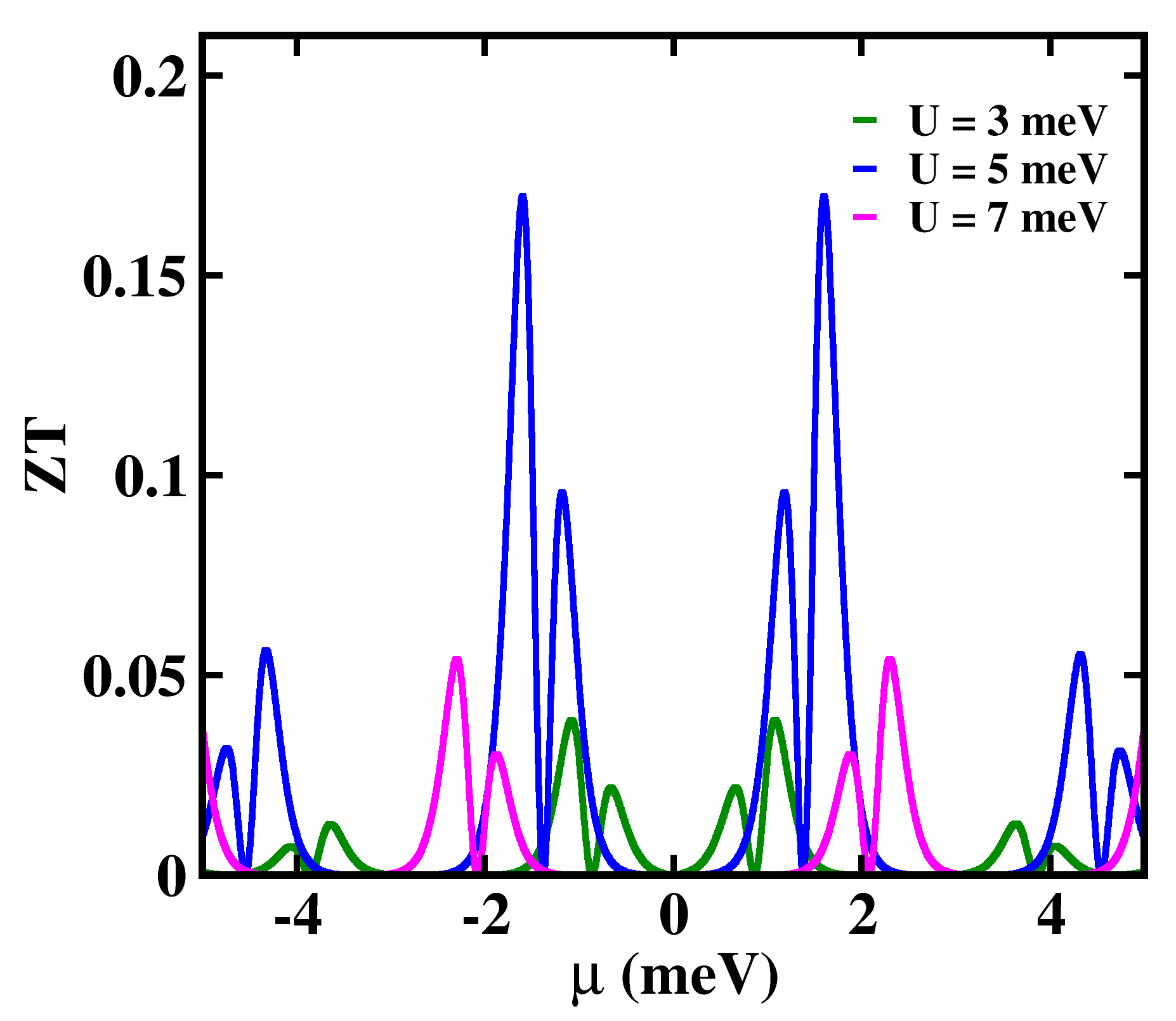}
	\caption{\label{all_U} Figure of merit ($ZT$) as a function of $\mu$ with different onsite Coulomb interaction ($U$) with $V=1~\text{meV}$, $t=1~\text{meV}$, $\Gamma=0.001~\text{meV}$ and magnetic flux $\Phi$= $\Phi_0$. }
\end{figure}

Another observation we have made is regarding the influence of the onsite Coulomb interaction ($U$) on the figure of merit for $\Phi=\Phi_{0}$, as depicted in Fig. \ref{all_U}. When the onsite Coulomb interaction $U$ increases from $0$ to $5 ~\text{meV}$, both the electrical conductance and $ZT$ increases. This behavior can be attributed to the interplay of Coulomb repulsion and quantum confinement. In a system with QDs, as the Coulomb interaction $U$ increases, it broadens the energy levels in QDs due to stronger electron-electron repulsion. This increased separation reduces the probability of multiple electrons occupying the same state, effectively broadening the energy levels. This broadening suppresses the bipolar effect, which enhances charge transport efficiency and improves thermoelectric performance, leading to increased electrical conductance and a higher $ZT$. Several theoretical studies on thermoelectric transport in QD based nanojunctions, without considering the Aharonov-Bohm effect, have reported similar trends \cite{trocha2012large,zheng2012thermoelectric,tagani2012thermoelectric,zimbovskaya2020thermoelectric}. Liu et al. observed that increasing the intra-dot Coulomb interaction $U$ in QDs broadens the energy level spacing, which weakens the bipolar effect, leading to a significant increase in $ZT$ \cite{liu2010enhancement}. Likewise, Zimbovskaya found that with an increase in $U$, both the Seebeck coefficient and $ZT$ are enhanced \cite{zimbovskaya2022large}.

Moreover, the wave function becomes more accommodating to various electron configurations without strict limitations on electron occupancy. This flexibility can result in more efficient charge carrier movement and better thermal management within the QD system. As a result, the $G_V$ of the system increases from $0.019~e^{2}/h$ ($U=0$) to $0.705~e^{2}/h$ ($U=5~\text{meV}$), as does the potential for the $S$ to rise (see the Fig. S3 in the supplementary information). Furthermore, we observe that, as U increases, the coherence between electronic states improves, leading to an enhancement in $ZT$. This increased coherence is evident from the increased value of corresponding transition rates (see Fig. S4(a) in the supplementary information) as transition rates in a sense are measure of coherence among the states. When $U=0$, the transition rates are relatively lower compared to the case when $U=5~\text{meV}$ (as shown in the Fig. S4(a), supplementary information). However, at very high Coulomb interaction strength ($U=7~\text{meV}$), electron-electron correlations become pronounced, leading to strong repulsive forces between electrons on the same site. This high repulsion decreases the overlap between the resultant states and existing relevant states. As a consequence, the transition rates are decreases for $U=7~\text{meV}$, as shown in Fig. S4(a), supplementary information. This reduction in transition rates results in a decrease in $ZT$ values for $U=7~\text{meV}$.
From Fig. S4 (b), we interestingly notice that, in the off-resonant condition, the probabilities $P_2$ and $P_3$ are getting farther away from each other as $U$ increases till $U=5~\text{meV}$ in contrast to the resonant condition where $P_2$ always equals to $P_3$.

\begin{figure}
	\centering
	\includegraphics[width=0.7\columnwidth]{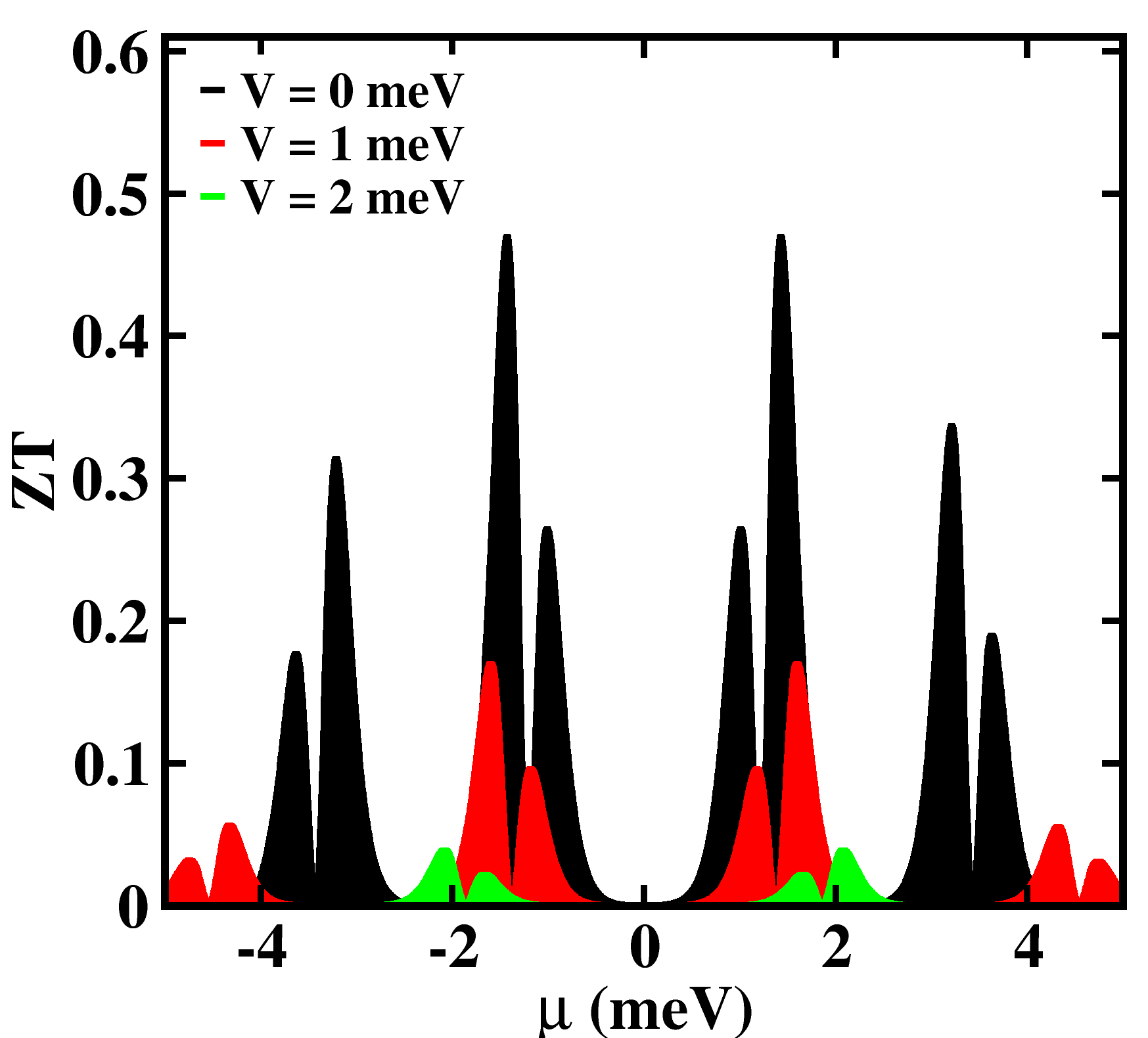}
	\caption{\label{all_V}Figure of merit ($ZT$) as a function of $\mu$ with different inter-site Coulomb interaction ($V$) with $U=5~\text{meV}$, $t=1~\text{meV}$, $\Gamma=0.001~\text{meV}$ and magnetic flux $\Phi$= $\Phi_0$.}
\end{figure}

Furthermore, tuning the inter-site Coulomb interaction ($V$) between electrons can significantly impact the transport properties of the QD system. Fig. \ref{all_V} illustrates the impact of the inter-site Coulomb interaction ($V$) on $ZT$ as a function of $\mu$. We observe that without $V$ the conductance exhibits a maximum value of $1.95~e^{2}/h$. This is due to the high transition matrix element values (see case-4 of section IV in supplementary information), indicating efficient electron transport, which leads to an increase in conductance. Thus, in the absence of $V$, the $ZT$ value exhibits a significant peak, reaching approximately $0.5$ for $\Phi = \Phi_{0}$, which is much larger than the value of state-of-the-art low-temperature thermoelectric materials. Zhou et al. reported a non-optimized $ZT$ value exceeding $0.55$ at $77~\text{K}$, highlighting the potential of strongly correlated quantum dot nanocomposites as high-efficiency thermoelectric materials at low temperatures \cite{zhou2010thermoelectric}. Without inter-site Coulomb repulsion, electron transport proceeds smoothly, encountering only the restrictions imposed by onsite Coulomb interaction. As a result, the system experiences good thermoelectric performance, leading to the observed significant peak in the $ZT$ value.

\subsubsection{Effect of dot-electrode coupling strength on ZT}

\begin{figure}
	\centering
	\includegraphics[width=1.0\columnwidth]{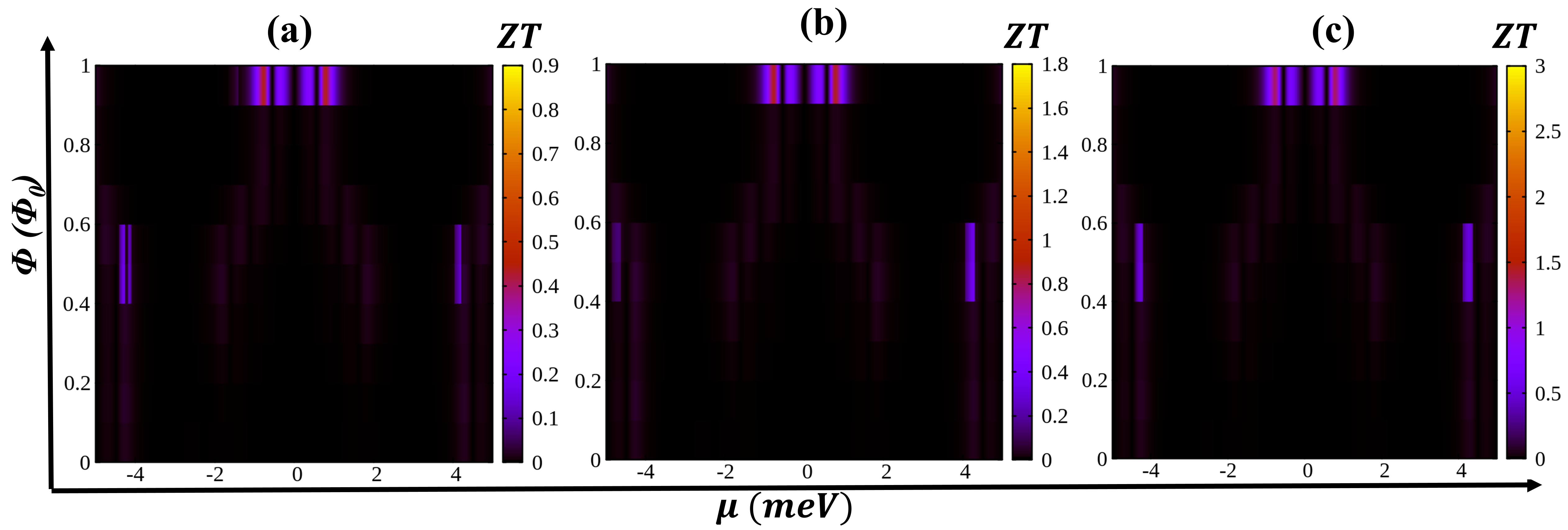}
	\caption{\label{diff_gamma} Figure-of-merit ($ZT$) as a function of $\mu$ and $\Phi$ at various dot-electrode coupling strength (a) $\Gamma$= $0.005~\text{meV}$, (b) $\Gamma$= $0.01~\text{meV}$, and (c) $\Gamma$= $0.015~\text{meV}$ with fixed $t=1~\text{meV}$, $U=5~\text{meV}$ and $V=1~\text{meV}$.}
\end{figure}

Fig. \ref{diff_gamma} illustrates the figure of merit for various dot-electrode coupling strengths ($\Gamma$) as a function of $\Phi$ and $\mu$ with $U = 5~\text{meV}$. Panels (a)-(c) correspond to different values of dot-electrode coupling strengths, denoted by $\Gamma$ = $0.005$, $0.01$, and $0.015~\text{meV}$, respectively. Notably, $ZT$ exhibits an increasing trend with the increase in $\Gamma$ ($ZT$ $\approx$ $0.9$, $1.8$, and $3$ when $\Gamma$ = $0.005$, $0.01$, and $0.015~\text{meV}$, respectively), indicating that stronger coupling between the triple QD and the electrodes results in larger $ZT$ values. This observation can be elucidated by considering the impact of coupling strength on charge carrier transport. 

Zheng et al. reported that in a single QD Aharonov-Bohm interferometer, weaker coupling between the dot and the electrodes results in a larger $ZT$ value \cite{zheng2012thermoelectric}. Similarly, Murphy et al. observed that a weaker coupling strength between the molecule and electrodes can lead to a violation of the Wiedemann-Franz law and an increase in $ZT$ \cite{murphy2008optimal}. However, in our study, we find the opposite trend: increasing the coupling strength between the QD and the electrodes leads to a higher $ZT$ value. As $\Gamma$ increases, the interaction between the QDs and electrodes strengthens, enhancing electron transmission through the QDs and reducing contact resistance. This results in higher electrical conductance and subsequently a higher $ZT$. 

This results align well with a recent theoretical and experimental study by Kleeorin et al. on the thermoelectric response of a single QD in a Coulomb blockade mesoscopic system. Their findings showed that the tunneling rate between the QD and the electrodes can be easily modified using split-gate technology. Moreover, they observed that increasing the coupling strength, $\Gamma$, enhances both conductance and thermopower, ultimately improving the thermoelectric performance \cite{kleeorin2019measure}. However, for our study, which focuses on the thermoelectric effects in the weak coupling or Coulomb blockade regime, we restrict our analysis to lower dot-electrode coupling strengths. This is crucial as it ensures the validity of the Pauli master equations used in our modeling, which are applicable in the weak coupling regime. It's essential to consider realistic coupling strengths that ensure the reliability of the results.

\subsection{Three QDs arranged asymmetrically in an AB ring}
\begin{figure}
	\centering
	\includegraphics[width=0.8\columnwidth]{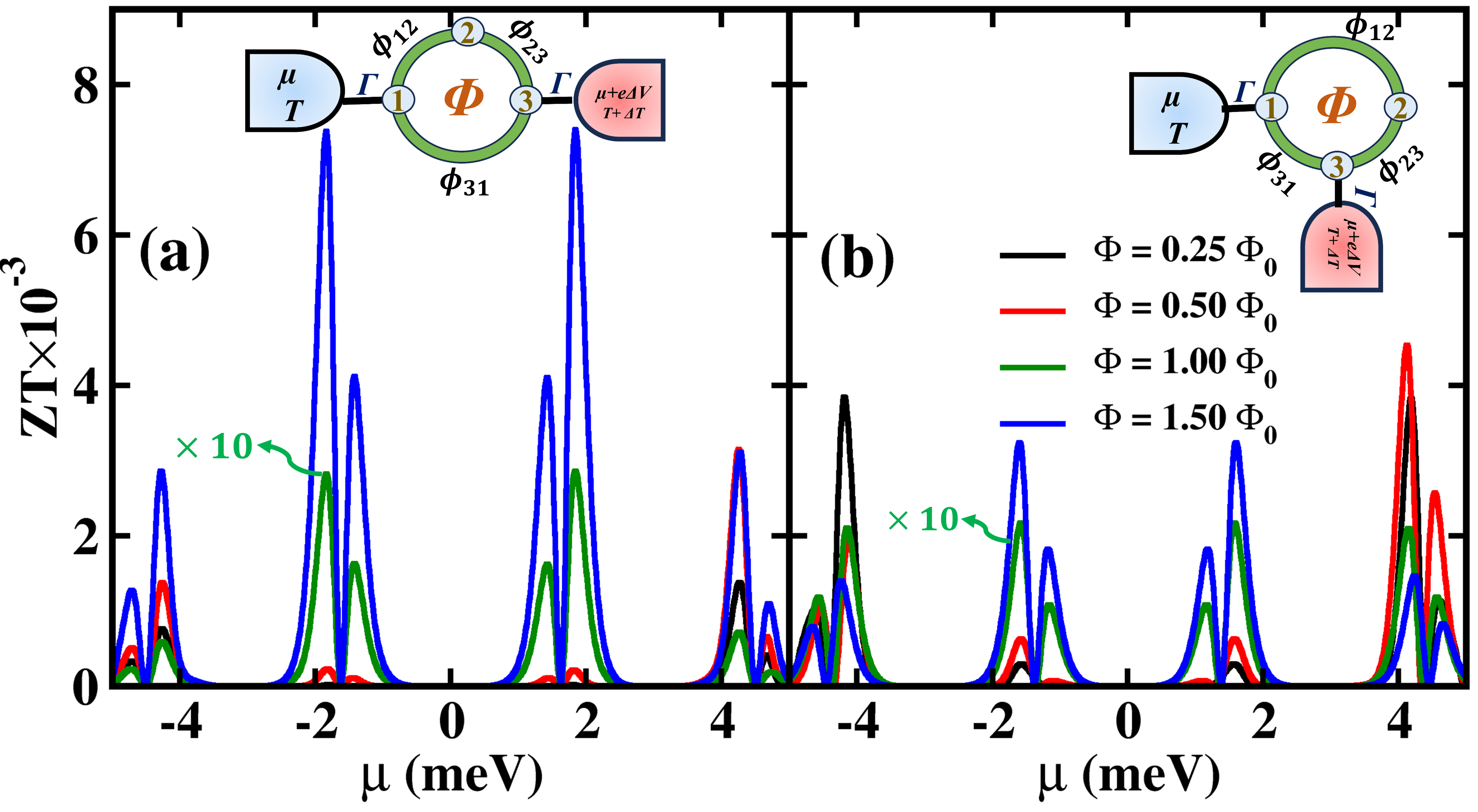}
	\caption{\label{asym_3_dot} Figure-of-merit ($ZT$) as a function of $\mu$ at different $\Phi$ with fixed $\Gamma=0.001~\text{meV}$, $U=5~\text{meV}$ and $V=1~\text{meV}$ for triple QDs asymmetrically embedded in an AB ring with two different connections (a)($\phi_{12}=\frac{\phi}{4},\phi_{23}=\frac{\phi}{4},\phi_{31}=\frac{\phi}{2}$) and (b) ($\phi_{12}=\frac{\phi}{2},\phi_{23}=\frac{\phi}{4},\phi_{31}=\frac{\phi}{4}$). The plot for $\Phi$=$\Phi_0$ (green line) has been scaled to multiples of 10 in $ZT$ values of both (a) and (b), respectively.}
\end{figure}

After delving into the thermoelectric transport properties of three QDs arranged symmetrically within an AB ring, our curiosity led us to investigate the consequences of creating asymmetric arrangements of triple QDs within the AB ring.
We consider two different asymmetric arrangments of three QDs in the AB ring. The AB phases in each arm for both the asymmetric configurations are given as: (a) ($\phi_{12}=\frac{\phi}{4},\phi_{23}=\frac{\phi}{4},\phi_{31}=\frac{\phi}{2}$), and (b) ($\phi_{12}=\frac{\phi}{2},\phi_{23}=\frac{\phi}{4},\phi_{31}=\frac{\phi}{4}$), which are illustrated in Fig. \ref{asym_3_dot} (inset). We note that, both asymmetric configurations (a) and (b) can be achieved by using same vector potential \(\vec{\mathbf{A}} = \left(-\frac{By}{2}, \frac{Bx}{2}, 0\right)\), used in the symmetric case (see the deatils in section I of supplementary information). The total AB phase around the entire loop remains the same as in the symmetric case ($\phi=\frac{2\pi\Phi}{\Phi_0}$). Although the same magnetic flux is applied in the asymmetric arrangement, we expect different physical results due to the altered configuration of the system itself. The motivation for considering these asymmetric geometries is to explore whether any other arrangements of QDs can yield a higher $ZT$ value compared to the symmetric one.

From our analysis, we observed that the maximum $ZT$ value observed in the asymmetric arrangements reached only around $0.0295$ (see Fig. \ref{asym_3_dot}). This value is remarkably lower in comparison to the  case of symmetric arrangement where $ZT$ is 0.15. Our results indicate that, asymmetric arrangement fails to produce $ZT$ values as significant as those observed in symmetric arrangements. This underscores the crucial role of symmetry in optimizing thermoelectric performance in QD systems and underscores the challenges associated with introducing asymmetry into such systems. 

\subsection{Four QDs embedded symmetrically in an AB ring}
After studying all thermoelectric coefficients in triple QDs system within an AB ring, we are interested to investigate the potential impact of increasing the number of QDs in the ring on thermoelectric applications. Exploring this aspect could provide valuable insights into the scalability and efficiency of such systems for thermoelectric device development. So, we have expanded our study to include a system with four QDs symmetrically embedded in an AB ring, as the symmetric arrangement proved superior for thermoelectric transport. Therefore, here we consider the AB phase for each arm which can be expressed as follows: $\phi_{12}=\phi_{23}=\phi_{34}=\phi_{41}=\frac{\phi}{4}=\frac{2\pi\Phi}{4\Phi_0}$. 

\begin{figure}
	\centering
	\includegraphics[width=0.9\columnwidth]{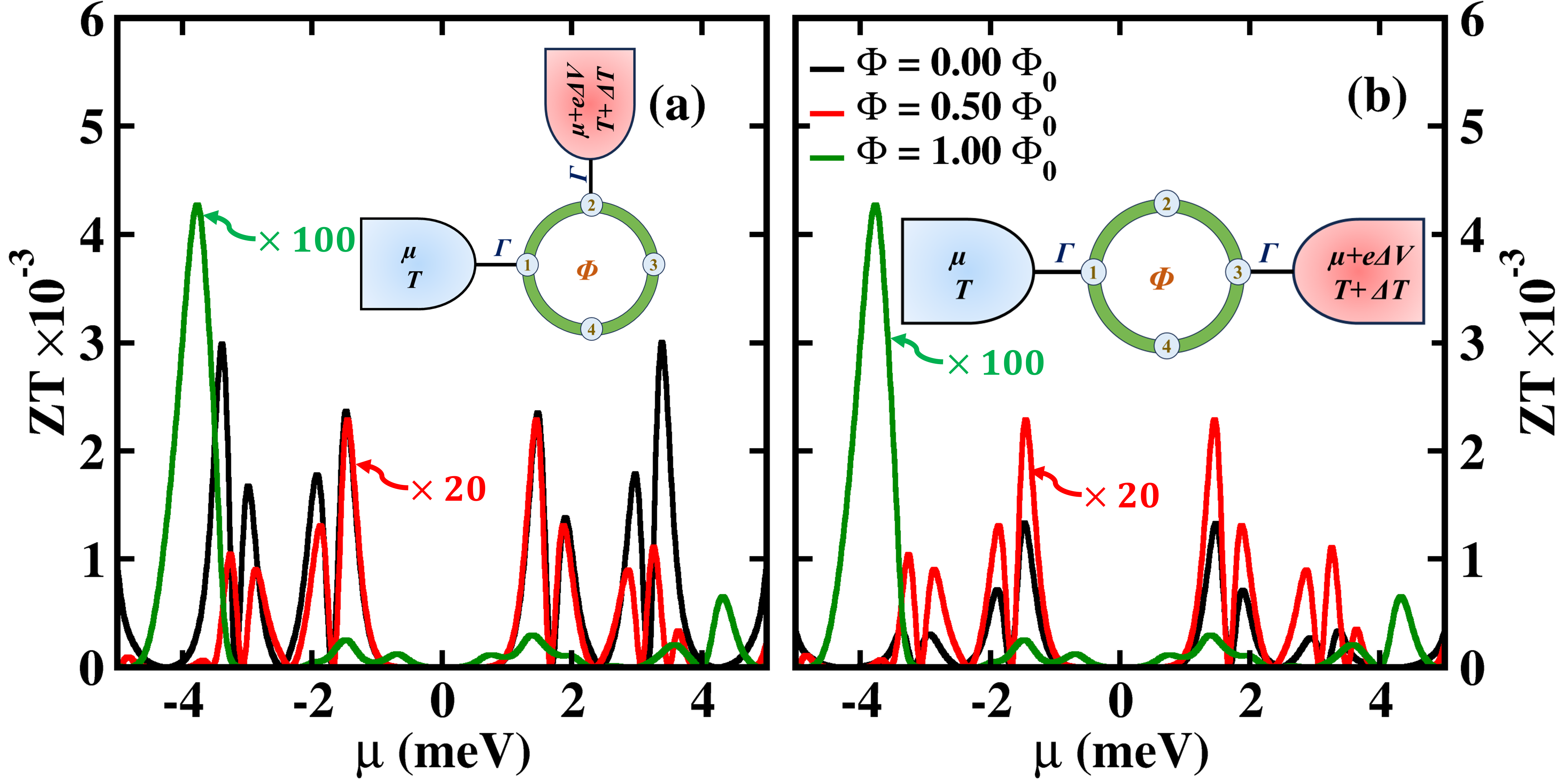}
	\caption{\label{4_dot} Figure of merit ($ZT$) as a function of $\mu$ with different $\Phi$ for four QDs symmetrically embedded in an AB ring. The onsite and inter-site Coulomb interactions is set $5~\text{meV}$ and $1~\text{meV}$, respectively, shown in two different connections: (a)consecutive connections (1-2 connection) and (b) alternating connections (1-3 connection). The plot for $\Phi$=0.50$\Phi_0$ and $\Phi$=$\Phi_0$ (red lines and green line) has been scaled to multiples of 20 and 100 in $ZT$ values, respectively.}
\end{figure}

The Fig. \ref{4_dot} (a) and (b) illustrates the variation of the $ZT$ as a function of $\mu$ at different $\Phi$ for nanojunctions containing four QDs embedded in an AB ring. Electrodes are connected in two different ways: one with consecutive connections (1-2 connection) and another with alternating connections (1-3 connection). The system includes onsite and inter-site Coulomb interactions of $5~\text{meV}$ and $1~\text{meV}$, respectively. Here we notice that, increasing the number of QDs embedded in an AB ring enhances the $ZT$ of the nanojunction, with a maximum $ZT$ value of approximately 0.43, for both connections as compared to triple QDs embedded in an AB ring. The enhancement of $ZT$ can be explained with several key factors. Firstly, additional QDs provide more pathways for charge carriers to traverse, effectively increasing the electrical conductance of the system. Secondly, the presence of multiple QDs can lead to quantum interference effects, where electrons traveling through different paths around the ring interfere constructively, enhancing the overall conductance and contributing to higher $ZT$ values. Moreover, with more QDs, the system can exhibit a greater number of resonant states, allowing for more efficient resonant tunneling of electrons through the nanojunction. This alignment of energy levels with the chemical potential further enhances conductance and facilitates the formation of sharp peaks in the conductance spectrum, ultimately leading to higher $ZT$ values. 

Overall, the increased number of QDs in the AB ring increases both the conductance and the quantum interference effects, resulting in improved thermoelectric performance and a higher $ZT$ which is beneficial for thermoelectric applications. This finding also aligns with previous theoretical studies, such as that by Zimbovskaya, who investigated the thermoelectric effect in multiple QDs arranged in a serial configuration and coupled to nonmagnetic conducting electrodes. Their study reported that the enhancement of both the Seebeck coefficient and $ZT$ becomes more pronounced with the increasing number of QDs, attributed to the presence of a Coulomb gap \cite{zimbovskaya2022large}.

\section{conclusions}

In conclusion, our investigation delves into the thermoelectric properties of three QDs embedded in an AB ring. We employ the Pauli master equation approach and linear response theory to explore these properties. These QDs are embedded symmetrically in the AB ring and are weakly coupled to metal electrodes. We observed a significant impact of magnetic flux on these properties, with quantum interference induced by the flux being a key factor affecting the thermoelectric effects. Without magnetic flux, the maximum conductance $G_V$ and figure of merit $ZT$ are approximately $2.68 \times 10^{-3}~e^{2}/h$ and $0.71 \times 10^{-3}$, respectively. However, as the magnetic flux increases from 0 to \(\Phi_0\), the value of \(G_V\) undergoes a remarkable 250-fold increase, reaching approximately \(0.705 \, e^2/h\). Meanwhile, the thermoelectric figure of merit \(ZT\) also sees a substantial 200-fold increase, reaching around \(0.15\). We also studied the effects of on-site and inter-site Coulomb interactions on $ZT$. Our observations indicate that increasing the on-site Coulomb interaction $U$ initially enhances $ZT$. However, at higher $U$ values ($U = 7~\text{meV}$), $ZT$ decreases due to strong electron-electron correlations. In the absence of inter-site Coulomb interactions $V$, $ZT$ can still increase up to 0.5. Therefore, to achieve an optimal $ZT$, a moderate $U$ and minimal $V$ are required.

Additionally, $ZT$ increases with dot-electrode coupling strength $\Gamma$. However, as our study focuses on the weak coupling regime, we limit our calculations to $\Gamma = 0.001~\text{meV}$. When comparing symmetric and asymmetric QDs arrangements in an AB ring, the symmetric arrangement exhibits superior thermoelectric performance. We also explored how increasing the number of QDs in the ring enhances the thermoelectric properties and found a high $ZT$ value of approximately $0.43$ (with inter-site Coulomb interaction) indicating the substantial improvement in thermoelectric performance as the number of QDs increases.

Many experimentalists have successfully studied AB oscillations and spin thermoelectric effects in QD rings separately. For example, Debbarma et al. found that one-dimensional quantum rings with two doubly connected QDs exhibit an AB period similar to that of higher-symmetry systems \cite{debbarma2021effects}. Modern fabrication techniques, such as those using PbTe heterostructures, enable the creation of AB rings with serially coupled PbTe-based QDs, which can exhibit effects like the spin Seebeck effect through the inverse spin Hall effect \cite{kobayashi2002tuning,springholz1993modulation}. Overall, our study highlights the substantial impact of magnetic flux on thermoelectric performance. It also demonstrates the potential of using multiple QDs arranged symmetrically in an AB ring, coupled with metal electrodes, for effective thermoelectric applications and advancements in device development.

\section*{References}
\bibliography{references.bib}
\bibliographystyle{unsrt}

\section*{Acknowledgement}
PS acknowledges DST-INSPIRE (IF190005) for financial support. PP thanks DST-SERB for ECRA project (ECR/2017/003305). 

\section*{Conflict of interest}
The authors declare that they have no conflicts of interest.

\section*{Data availability}
The datasets used and/or analyzed during the current study available from the corresponding author on reasonable request.

\end{document}